\documentstyle[12pt]{article}
\begin{document}
\date{\today}
%
\def\R{\rm l\!R\,}
%
\def\N{\rm l\!N}
%
\def\Compton{\overline{\phantom{\overline{.\!.}}}\!\!\!\lambda}
%
%
%
%
\title{\bf Fermionic Casimir effect in an external magnetic
field\footnote{Talk presented by A C Tort at the {\it IV workshop on quantum
field theory under the influence of external conditions}}}
\author{M. V. Cougo-Pinto\thanks{e-mail: marcus@if.ufrj.br}, 
C. Farina\thanks{e-mail: farina@if.ufrj.br},
A. Tort\thanks{e-mail: tort@if.ufrj.br}\\
\\Instituto de F\'{\i}sica, Universidade Federal do Rio de 
Janeiro\\
CP 68528, Rio de Janeiro, RJ 21945-970, Brazil\\
\\}
\maketitle
\begin{abstract}
The influence of an external constant uniform magnetic field on the Casimir
energy density of a Dirac field under antiperiodic (and periodic) boundary
condition is computed by applying Schwinger's proper time method. The result
thus obtained shows that in principle, under suitable conditions, the
magnetic field can enhance the fermionic Casimir energy density.
\end{abstract}
%
%
\section*{Introduction}
The Casimir effect \cite{Casimir48} can be generally defined as the effect
of non-trivial space topology on the vacuum fluctuations of relativistic
quantum fields \cite{MostepanenkoTrunovB,PlunienMullerGreinerR}. The
corresponding change in the vacuum fluctuations appears as a shift in the
vacuum energy density and a resulting vacuum pressure. In the case of the
electromagnetic Casimir effect, there are three experiments involving
metallic surfaces \cite{CasExp}. The results, particularly the two more
recent ones, are in accord with theoretical predictions.

The Casimir effect has been computed for fields other than the
electromagnetic and boundary conditions different from the one implemented
by conducting surfaces. The fermionic Casimir effect is of particular
importance due, {\it e.g.}, to the fundamental role played by the electron
in QED and the quarks in QCD; it was first computed by Johnson
\cite{Johnson75} for applications in the MIT bag model \cite{MITbagmodel},
in which the Casimir energy density is an important ingredient. For a
massless Dirac particle Johnson's result predicts an energy density $7/4$
times the energy density of the electromagnetic Casimir effect. In the case
of the Casimir effect of an electrically charged quantum field it is natural
and important to ask how an external electromagnetic field may influence the
fluctuations and, consequently, the effect. Indeed, we should expect on
physical grounds the existence of such an influence and it is necessary to
calculate its features and magnitude in order to clarify its role and to
obtain a deeper understanding of the Casimir effect. Even if the conclusion
turns out to be that the magnitude of the effect is negligible in realistic
situations, nonetheless it can be argued that similarly to other vaccum
effects, such as the Scharnhorst effect \cite{Scharnhorst-Barton}, it is a
matter of first principles and as such it must be investigated. Here we
propose to investigate the influence of an external field by considering a
Dirac quantum field under antiperiodic (and perodic) boundary conditions
\cite{Ford80}. These choices of geometry and quantum and external fields
avoid technicalities in the formalism and will permit us to focus our
attention on the fundamental issue, which is the physical effect of the
external field on the fermionic vacuum energy density. Once this influence
is understood the path is open to consider more complicated geometries and
external fields as well as other quantum vacua.
%
\section*{Fermionic Casimir energy density for antiperiodic boundary conditions}
In order to accomplish the investigation which we have described above we
will employ the method proposed by Schwinger which stems originally from his
source theory  \cite{Schwinger92}. Since the method has been clearly
explained by Schwinger \cite{Schwinger92} and already applied to several
situations \cite{CPFSST,CPFkappa} we will recall briefly its main features.
The vacuum energy is given by
\begin{equation}
{\cal E}_0=-{{\cal W}^{(1)}\over T}\;,
\end{equation}
where ${\cal W}^{(1)}$ is the one-loop effective action and $T$ is the
duration of the measurement. In the case of a fermionic field Schwinger's
proper time formula for the effective action is given by
\begin{equation}\label{W}
{\cal W}^{(1)}={i\over 2}\int_{s_o}^\infty
{ds\over s}\;Tr\,e^{-isH}\; ,
\end{equation}
where $s_o$ is a cutoff in the proper time $s$, $Tr$ means the total trace
and $H$ is the proper time Hamiltonian, which for charged fermions in an
external electromagnetic fiels reads
\begin{equation}
H=(P-eA)^2-(e/2)\sigma_{\mu\nu} F^{\mu\nu}+m^2\;,
\end{equation}
where $P$ has components $P_\mu=-i\partial_\mu$, $e$ is the charge of the
Dirac particle, $A$ is the electromagnetic potential, $F$ is the
electromagnetic field, which here appears contracted with the combination of
gamma matrices $\sigma_{\mu\nu}=$$i[\gamma_\mu,\gamma_\nu]/2$, and $m$ is
the mass of the Dirac particle. We will consider first the influence of a
uniform external magnetic field in the case of antiperiodic boundary
condition on the ${\cal OZ}$-direction.  The imposition of antiperiodic
boundary conditions gives for the z-component of $P$ the eingenvalues $\pi
n/a$, where $n$ runs in the set of odd integers. The other space components
of $p$ are constrained into the Landau levels created by the magnetic field
${\bf B}$; we call $B$ the component of ${\bf B}$ along the ${\cal OZ}$-axis
and for convenience the axis positive direction is chosen in such a way that
$eB$ is positive. The trace in (\ref{W}) is then given by:
\begin{eqnarray}\label{Tr}
Tr\,e^{-isH}=e^{-ism^2}\sum_{\alpha=\pm 1}2 
\sum_{n=-\infty}^{\infty}e^{-is[\pi (2n+1)/a]^2}
\sum_{n^\prime=0}^{\infty}{eB{\cal A}\over
2\pi}e^{-iseB(2n^\prime+1-\alpha)}
\int{dt\,d\omega\over 2\pi}e^{is\omega^2}\; ,
\end{eqnarray}
where the first sum takes care of the four components of the Dirac spinor,
the second sum is over the eigenvalues stemming from the antiperiodic
boundary condition, the third sum is over the Landau levels with the
corresponding multiplicity 
factor due to  degeneracy, and the integral range is given by the 
measurement time $T$ and by the continuum of eingenvalues $\omega$ of the
operator $P^o$. Following Schwinger we apply Poisson sum formula
\cite{Poisson1823} to the second sum and sum straightforwardly over the
Landau levels to recast the trace into the following form:
\begin{eqnarray}\label{Trfinal}
Tr\,e^{-isH}={T\,a{\cal A}\over 4\pi^2i}{e^{-ism^2}\over s^2}
\left[1+2\sum_{n=1}^{\infty}(-1)^n
e^{i(an)^2/4s}\right]\left[1+iseB\,L(iseB)\right]\;,
\end{eqnarray}
where $L(\xi)=\coth \xi-\xi^{-1}$ is the Langevin function. Taking
(\ref{Trfinal}) into (\ref{W}) we obtain for the effective action:
\begin{equation}\label{Wfinal}
{\cal W}^{(1)}={\cal L}^{(1)}(B)\,Ta{\cal A}-\rho_{\mbox{\tiny
AP}}^{\mbox{\tiny total}} (a,B)\,a{\cal A}T\; ,
\end{equation}
where on the r.h.s. the first term gives the (unrenormalized)
Heisenberg-Euler effective Lagrangian \cite{H-E}:
\begin{equation}\label{H-E}
{\cal L}^{(1)}(B)={1\over 8\pi^2}\int_{s_o}^\infty {ds\over s^3}
e^{-ism^2}\,iseB\,\coth(iseB)\; ,
\end{equation}
and the second term gives the (still cutoff-dependent) Casimir energy density
\begin{eqnarray}\label{E(a,B,s0)}
\rho_{\mbox{\tiny AP}}^{\mbox{\tiny total}} (a,B)=-{1\over
4\pi^2}\sum_{n=1}^{\infty}(-1)^n
\int_{s_o}^\infty{ds\over s^3}e^{-ism^2+i(an)^2/4s}\;
\left[1+iseB\,L(iseB)\right]\; ,
\end{eqnarray}
which is the quantity we are interested in. The Heisenberg-Euler 
Lagrangian makes no contribution to the Casimir energy density, since it
exhibits no dependence on the separation $a$. Therefore it does not take
part of the expression for the Casimir energy density, which is set to zero
when $a\to\infty$. We now make the change of the integration variable to
$\sigma$$=a^2/is$. The limit in which the cutoff $s_o$ goes to zero can be
taken and in the resulting expression the part of the Casimir energy density
which exists in the absence of the external magnetic field can be expressed
in terms of the modified Bessel function $K_2$ (formula {\bf 3.471},9 in
\cite{Grad}). 
The resulting expression for the total vacuum energy density is:
\begin{eqnarray}\label{E(a,B)}
\rho_{\mbox{\tiny AP}}^{\mbox{\tiny total}} (a,B) &=&
 2{(am)^2\over \pi^2 a^4}\sum_{n=1}^{\infty}
{(-1)^{n}\over n^2}\,K_2(amn)\nonumber\\
&+&{eB\over 4\pi^2a^2}\sum_{n=1}^{\infty}(-1)^{n}
\int_0^\infty d\sigma
e^{-n^2\sigma/4-(am)^2/\sigma}\;
L(eBa^2/\sigma)\; .
\end{eqnarray}
When there is no external magnetic field $B$ the Casimir
energy density is given by the first term on the r.h.s. of equation
(\ref{E(a,B)})
\begin{equation}\label{E(a,0)}
\rho_{\mbox{\tiny AP}}(a,0)=
2{(am)^2\over \pi^2 a^4}\sum_{n=1}^{\infty}
{(-1)^{n}\over n^2}\,K_2(amn)\; .
\end{equation}
Here we are interested in the second term on the r.h.s. of equation
(\ref{E(a,B)}), which measures the influence of the external magnetic field
in the Casimir energy density. The contribution of the magnetic field is
given by a quadrature, which is strictly positive, decreases monotonically
as $n$ increases and goes to zero in the limit $n\rightarrow\infty$.
Consequently, we have by Leibnitz criterion a convergent alternating series
in (\ref{E(a,B)}) and we may conclude that the external magnetic field
increases the fermionic Casimir energy density. This is the main result of
this work, which elucidates part of the interplay between two of the most
fundamental phenomena in relativistic quantum field theory, namely: the
Casimir effect and the fermionic vacuum properties described by the
Euler-Heisenberg effective Lagrangian. Remark that the same fermionic vacuum
acted by an external field under antiperiodic boundary condition can be
studied from a different point of view, in which we look for changes in the
vacuum constitutive relations due to confinement. This point of view
requires a different analysis and leads to quite different physical
phenomena\cite{CPFRT}.
%
\section*{Strong magnetic field regime for antiperiodic boundary conditions}
Consider the strong field regime in which changes in the charged vacuum may
be easier to occur \cite{GreinerMR}. The integral in equation (\ref{E(a,B)})
is dominated by the exponential function, whose maximum is $\exp(-amn)$ and
occurs at $\sigma$$=2am/n$. Therefore, in the strong field regime we are
justified in substituting the Langevin function by $1-\xi^{-1}$, which in
the cases $am\ll 1$ and $am\gg 1$ is characterised, respectively, by $\vert
B\vert \gg \vert\phi_o\vert/a^2$ and 
$\vert B\vert\gg (\vert\phi_o\vert/a^2)(a/\Compton_c)$, where $\phi_o$ is
the fundamental flux $1/e$ and $\Compton_c$ is the Compton wavelength $1/m$.
For antiperiodic boundary conditions in the strong field regime, in both
cases, the second term in (\ref{E(a,B)}) can also be expressed in terms of a
modified Bessel function (formula {\bf 3.471},9 in \cite{Grad}), and the
Casimir energy density can be written as:
\begin{equation}\label{E(B>>)}
\rho_{\mbox{\tiny AP}} (a,B)=
{eBm\over \pi^2a^2}(am)^2\sum_{n=1}^{\infty}{(-1)^{n}\over n}\,K_1(amn)\;.
\end{equation}
\par
In the limit $am\ll 1$, after discarding second order terms in the 
expansions of the corresponding Bessel functions (formula {\bf 8.446} in
\cite{Grad}), we obtain from (\ref{E(a,0)}) and (\ref{E(B>>)}):
\begin{equation}\label{E(am<<B=0)}
\rho_{\mbox{\tiny AP}} (a,0)=-7{\pi^2\over 180 a^4}\hskip
0.5 cm (am\ll 1)\; ,
\end{equation}
which is essentially the result obtained in \cite{Ford80} and
\begin{equation}\label{E(am<<B>>)}
\rho_{\mbox{\tiny AP}} (a,B)=- {eB\over 12\;a^2}\hskip 0.5 cm 
(am\ll 1,\; \vert B\vert\gg \vert\phi_o\vert/a^2)\;.
\end{equation}
From (\ref{E(am<<B=0)}) and (\ref{E(am<<B>>)}) we can see that the ratio
between the Casimir energy density with the strong external magnetic field
and the Casimir
energy without this field is
\begin{equation}\label{Ratio(am<<)}
{\rho_{\mbox{\tiny AP}} (a,B)\over\rho_{\mbox{\tiny AP}} (a,0)}={15\over
7\pi^2}{B\over
\phi_o/a^2}\hskip 0.5 cm 
(am\ll 1,\; \vert B\vert\gg \vert\phi_o\vert/a^2)\;.
\end{equation}
A rough numerical estimation can be produced. If we set $a=1\mu$m we find
that $\rho_{\mbox{\tiny AP}} (a,B)\approx\rho_{\mbox{\tiny AP}} (a,0) \times
10^{-4}\;B/Tesla$.
We now examine the opposite limit, namely $am\gg 1$. In this limit 
we can use the assymptotic expansion of the corresponding Bessel 
functions (formula {\bf 8.451},6 in \cite{Grad}) to obtain from
(\ref{E(a,0)}) and (\ref{E(B>>)}):
\begin{equation}\label{E(am>>B=0)}
\rho_{\mbox{\tiny AP}} (a,0)=-4\left({am\over
2\pi^3}\right)^{3/2}\;{e^{-am}\over a^4}
\hskip 0.5 cm (am\gg 1)\; ,
\end{equation}
and
\begin{eqnarray}\label{E(am>>B>>)}
\rho_{\mbox{\tiny AP}} (a,B)=-{eB\over a^2}\left({am\over 2\pi^3}\right)^{1/2}
\;e^{-am}\hskip 0.5 cm (am\gg 1,\; \vert B\vert\gg
(\vert\phi_o\vert/a^2)(a/\Compton_c)\,)\;.
\end{eqnarray}
and from (\ref{E(am>>B=0)}) and (\ref{E(am>>B>>)}):
\begin{eqnarray}\label{Ratio(am>>)}
{\rho_{\mbox{\tiny AP}} (a,B)\over \rho_{\mbox{\tiny AP}} (a,0)}={B\over
(2\phi_o/a^2)(a/\Compton_c)}\hskip 0.5 cm (am\gg 1,\; \vert B\vert\gg
(\vert\phi_o\vert/a^2)(a/\Compton_c)\,)\;.
\end{eqnarray}
We can also try a rough numerical estimation for this case. For $a=1\mu$m we
have $\rho_{\mbox{\tiny AP}} (a,B) \approx 10^{-10} \rho_{\mbox{\tiny AP}}
(a,0) B/Tesla$.
%
\section*{\bf Final remarks and conclusion}
Results for the periodic case can also be obtained in the same way, here we
will only give the main result.  The total Casimir energy density for
periodic boundary conditions is given by
\begin{eqnarray}
\rho_{\mbox{\tiny A}} (a,B) &=&
2{(am)^2\over \pi^2 a^4}\sum_{n=1}^{\infty}
{1\over n^2}\,K_2(amn)\nonumber\\
&+&{eB\over 4\pi^2a^2}\sum_{n=1}^{\infty}
\int_0^\infty d\sigma
e^{-n^2\sigma/4-(am)^2/\sigma}\;
L(eBa^2/\sigma)\; .
\end{eqnarray}
The first term is the Casimir energy density in the absence of the external
magnetic field, the second one takes into account the presence of this
field. Notice that the main difference between the antiperiodic and periodic
results is the absence of the factor $(-1)^n$ in the summations. As in the
antiperiodc case, we can investigate the strong field regime (and also the
weak field regime) and repeat the analysis we did for the antiperiodic case.
These results will be published elsewhere.

Summing up, we have obtained the general expression of the fermionic Casimir
energy density under the effect of an external magnetic field for
antiperiodic and periodic boundary conditions. The results show that the
external field increases the Casimir energy density and reveals the
interplay between two agents which are known to affect vacuum fluctuations,
namely: external fields and non-euclidean space topology. We have derived an
expression for the vacuum energy density in the regime of strong magnetic
field and in this regime we have also obtained the small and large mass
limits of this energy density. Our formalism has a natural extension to more
complicated gauge groups and consequently may be useful in the investigation
of the QCD vacuum; this will be the subject of forthcoming work. As a final
commnent we observe that it may be also interesting to investigate the
combined action of confinement and applied magnetic field in the analogues
of Casimir effect which occurs in condensed matter physics and critical systems
\cite{Krech}.

\section*{\bf Acknowledgements}
A. C. Tort and C. Farina would like to acknowledge Prof. Dr. Michael Bordag
(Universit\"{a}t Leipzig), the Organizing Commitee of the {\it IV Workshop
on Quantum Field Theory under the influence of external conditions} and the
Brazilian governamental agency CAPES (Coordena\c c\~ao de Aperfei\c coamento
de Pessoal de Ensino Superior) for the hospitality and financial support
which made possible for both to be in Leipzig. A. C. Tort would like also to
acknowledge FAPERJ (Funda\c c\~ao de Amparo \`a Ci\^encia do Estado do Rio
de Janeiro) for partial financial support. C. Farina and M. V. Cougo-Pinto
would like to acknowledge also CNPq (Conselho Nacional de Pesquisas) for
partial financial support.

\begin{thebibliography}{99}
%
\bibitem{Casimir48} Casimir H B G 1948 {\it Proc. Kon. Nederl. Akad.
Wetensch.} {\bf 51} 793 
%
\bibitem{MostepanenkoTrunovB} Mostepanenko V M and Trunov N N 1997  
{\it The Casimir Effect and its Applications} (Oxford: Clarendon);
Mostepanenko V M and Trunov N N 1988 {\it Sov. Phys. Usp.} {\bf 31} 965
%
\bibitem{PlunienMullerGreinerR} Plunien G, Muller B and Greiner W 1986 {\it
Phys. Rep.} {\bf 134} 89
%
\bibitem{CasExp} Sparnaay M J 1958 {\it Physica} {\bf 24} 751; Lamoreaux  S
K 1997 {\it Phys. Lett.} {\bf 78} 5; Mohideen U and Roy A: {\it A precision
measurement of the Casimir force from $0.1$ to $0.9$ $\mu$m}. hep-ph/9805038
%
\bibitem{Johnson75} Johnson K 1975 {\it Acta Phys. Polonica} {\bf B6} 865
%
\bibitem{MITbagmodel} Chodos A, Jaffe R L, Johnson K, Thorn C B and
Weisskopf V F, 1974 {\it Phys. Rev.} {\bf D9} 3471; Chodos A, Jaffe R L,
Johnson K and Thorn C B 1974 {\it ibid.} {\bf D10} 2559; DeGrand T, Jaffe R
L, Johnson K and Kiskis J 1975 {\it ibid.} {\bf D12} 2060; Donoghue J F,
Golowich E and Holstein B R 1975 {\it ibid.} {\bf D12} 2875
%
\bibitem{Scharnhorst-Barton} Scharnhorst K 1990 {\it Phys. Lett.} {\bf B236}
354; Barton G 1990 {\it Phys. Lett.} {\bf B237} 559
%
%
\bibitem{Ford80} Ford L H 1980 {\it Phys. Rev.} {\bf D21} 933
%
\bibitem{Schwinger92} Schwinger J 1951 {\it Phys. Rev.} {\bf 82} 664;
Schwinger J 1992 {\it Lett. Math. Phys.} {\bf 24} 59
%
\bibitem{CPFSST} Cougo-Pinto M V, Farina C and S\'egui-Santonja A J
1994 {\it Lett. Math. Phys.} {\bf 30} 169; {\bf 31} 309. 
Cougo-Pinto M V, Farina C and Tort A 1996 {\it Lett. Math. Phys.} {\bf 37} 159
%
\bibitem{CPFkappa} Cougo-Pinto M V and Farina C 1997 {\it Phys. Lett.} {\bf
B391} 67
%
\bibitem{Poisson1823} Poisson S D 1823 {\it Journal de l'Ecole
Polytechnique} {\bf XII}, (cahier XIX) 420
%
\bibitem{H-E} Heisenberg W, 1935 {\it Z. Phys.} {\bf 90} 209; 
Euler H and B. Kockel B 1935 {\it Naturwissensch.} {\bf 23} 246;
Heisenberg W and Euler H 1936 {\it Z. Phys.} {\bf 98} 714;
Weisskopf V S 1936 {\it K. Dan. Vidensk. Selsk. Mat. Fys. Medd.} {\bf 14},
no. 6 reprinted in Schwinger J {\it Quantum Electrodynamics} (Dover, New
York,1958). For an English translation of Weisskopf's paper see Miller A I
{\it Early Quantum Electrodynamics} (CUP, Cambridge GB 1994)
%
\bibitem{Grad} Gradshteijn I S and Ryzhik I M {\it Tables of
Integrals, Series, and Products}, Academic Press, New York, London, 1965
%
\bibitem{CPFRT} Cougo-Pinto M V, Farina C, Rafelski J and Tort A
{\it Magnetic permeability of confined fermionic vacuum}, hep-th/9711190 v3
(To appear in Phys. Lett. {\bf B})
%
\bibitem{GreinerMR} Greiner W, M\"uller B and Rafelski J {\it Quantum
Electrodynamics of Strong Fields} (Springer-Verlag, Berlin, 1985)
%
{\bf B211} 1
%
%
\bibitem{Krech} Krech M {\it The Casimir Effect in Critical System} (World
Scientific, Singapore, 1994)
%
\end{thebibliography}
\end{document}